\documentclass[%
amsmath,amssymb,
aps,
pre,
twocolumn,
superscriptaddress
]{revtex4-1}

\usepackage{amsmath,amsfonts}
\usepackage{ae,aecompl}   
\usepackage{soul}
\usepackage{graphics}
\usepackage{graphicx}
\usepackage{dcolumn}
\usepackage{bm}
\usepackage{systeme}

\usepackage{geometry}
\usepackage{url}
\usepackage{subfigure}
\usepackage[english]{babel}
\usepackage[utf8x]{inputenc}
\usepackage[T1]{fontenc}

\usepackage[usenames,dvipsnames]{xcolor}

\usepackage{color}
\definecolor{darkgreen}{rgb}{0,0.6,0}
\definecolor{darkblue}{rgb}{0,0,0.6}
\definecolor{darkred}{rgb}{0.6,0,0}
\definecolor{darkpurple}{rgb}{0.5,0,0.5}
\newcommand{\nccom}[1]{\textcolor{black}{#1}}
\newcommand{\F}{\mathcal{F}}
\newcommand{\revtwo}[1]{\textcolor{black}{#1}}
\newcommand{\revthree}[1]{\textcolor{black}{#1}}
\newcommand{\revfour}[1]{\textcolor{black}{#1}}

\usepackage{hyperref}

\hypersetup{
	bookmarksopen=true,
	pdftitle=,
	pdfauthor=, 
	pdftoolbar=false,           
	pdfstartview={FitH},		
	pdfmenubar=true,			
	pdfhighlight=/O,			
	colorlinks=true,			
	urlcolor=darkblue,
	citecolor=darkblue,		    
	linkcolor=darkpurple	    
}

\begin{document}

\preprint{APS/123-QED}

\title{Phase separation on surfaces in presence of matter exchange}

\author{Nirvana Caballero}
\email[Corresponding author: ]{Nirvana.Caballero@unige.ch}
\affiliation{Department of Quantum Matter Physics, University of Geneva, 24 Quai Ernest-Ansermet, CH-1211 Geneva, Switzerland}

\author{Karsten Kruse}
\affiliation{Department of Biochemistry, University of Geneva, 1211 Geneva, Switzerland}
\affiliation{Department of Theoretical Physics, University of Geneva, 1211 Geneva, Switzerland}
\affiliation{NCCR Chemical Biology, University of Geneva, 1211 Geneva, Switzerland}

\author{Thierry Giamarchi}
\affiliation{Department of Quantum Matter Physics, University of Geneva, 24 Quai Ernest-Ansermet, CH-1211 Geneva, Switzerland}

\date{\today}
\begin{abstract}
We present a field theory to describe the composition of a surface spontaneously exchanging matter with its bulk environment. By only assuming matter conservation in the system, we show with extensive numerical simulations that, depending on the matter exchange rates, a complex patterned composition distribution emerges on the surface. For one-dimensional systems we show analytically and numerically that coarsening is arrested and as a consequence domains have a characteristic length scale. Our results show that the causes of heterogeneous lipid composition in cellular membranes may be justified in simple physical terms.
\end{abstract}

\maketitle

Living cells are full of fluid lipid membranes~\cite{alberts2008}. The primary function of these membranes is to compartmentalize the cell interior and to separate the cell from its environment. At the same time, diverse patterns that play essential roles in vital processes form on their surfaces. For example, protein clusters acting as units for sensing extra- or intracellular signals~\cite{Maddock:1993tu,varma1998,plowman2005,Fujita2007}. These protein clusters can be transient or not and are often associated with domains rich in specific kinds of lipids, commonly designated as lipid rafts~\cite{simons1997}. Another spectacular example of membrane-associated patterns are protein waves~\cite{Beta:2017gp}. Such waves can be standing~\cite{raskin1999} or traveling~\cite{Vicker:2000wj}, which can lead to turbulent dynamics~\cite{tan2020}. Some of the surface-associated patterns could be reproduced in reconstitution experiments in vitro~\cite{loose2008,Landino2021}. 

The physical principles underlying the formation of these patterns are still not fully understood and simple reaction-diffusion systems as pioneered by Turing~\cite{Turing:1952ja} can miss essential aspects. For example, convective transport along the membrane surface can play an important role~\cite{fan2008}. This holds notably for patterns associated with the so-called cytoskeleton, a cellular polymer network in which chemical energy is transformed into mechanical stress~\cite{alberts2008}. Gradients in this stress lead to flows along the membrane surface~\cite{Bois:2011kx,Kumar:2014fx,Hannezo:2015ba}.

\begin{figure}
\begin{center}
{\includegraphics[width=1\linewidth]{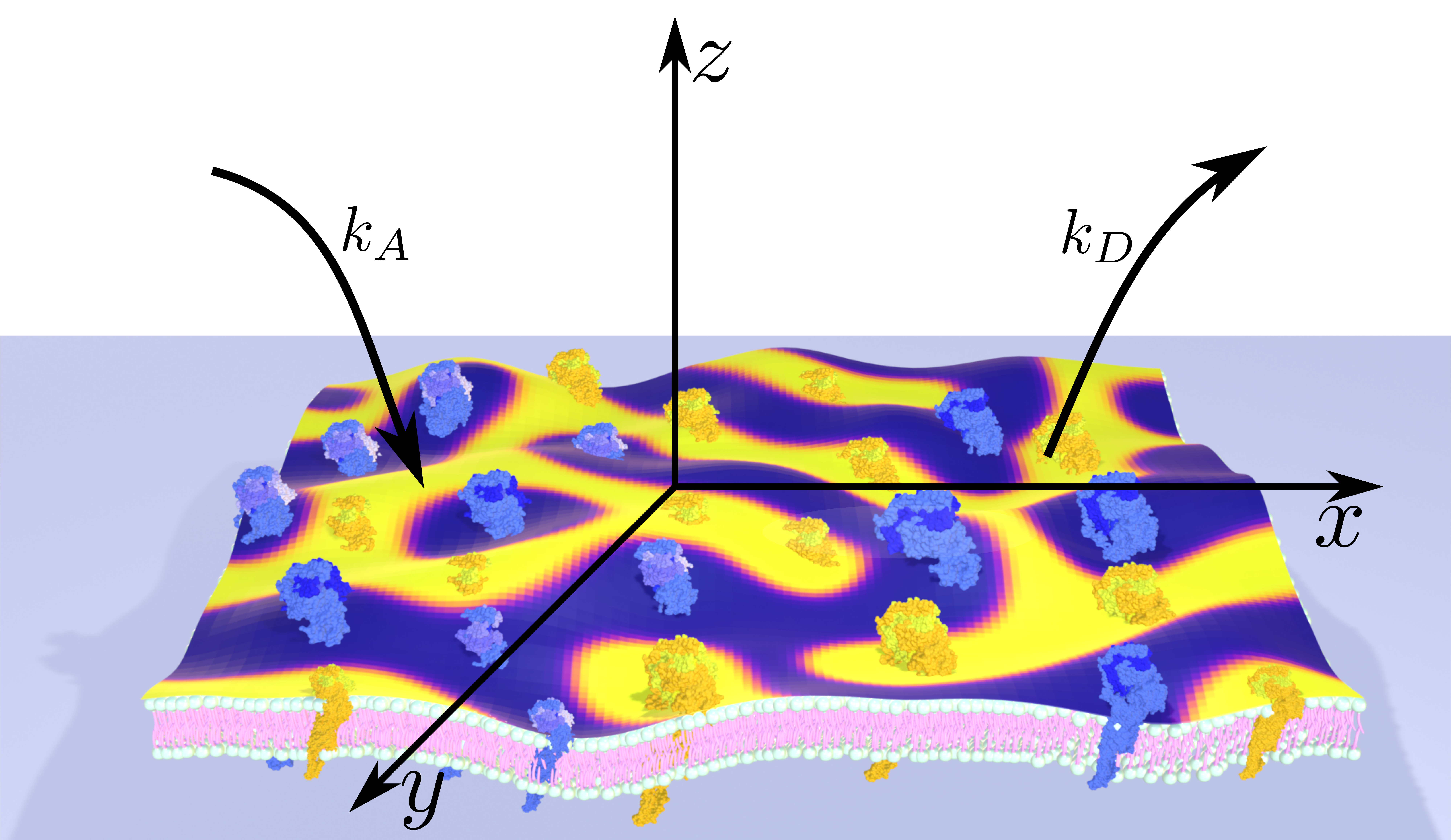}}
\end{center}
\caption{Illustration of the dynamics considered. Two types of particles (blue and yellow) on a membrane phase, in equilibrium with their respective bulk reservoirs, phase separate into blue and yellow domains. Each species attaches to and detaches from the membrane with the respective rates $k_A$ and $k_D$.}
  \label{fig:scheme}
\end{figure}

Alternatively, in-plane rearrangements in or on a membrane surface can result from phase separation. This is of particular importance for the formation of lipid domains. Lipid phase separation usually results in complete de-mixing, although the coupling between line tension at the interface between different phases and membrane bending can lead to stable domain patterns~\cite{baumgart2003}. Alternatively, membrane-associated patterns can be formed when the system is kept out of thermodynamic equilibrium~\cite{fang2019}, for example, through the exchange of matter between the surface and the surrounding medium~\cite{foret2005simple,turner2005} as shown schematically on  Fig.~\ref{fig:scheme}. Such an exchange is also essential for the formation of some protein patterns~\cite{Wettmann:2018bp}.

Pattern formation in presence of matter exchange between a surface and the surroundings is often studied theoretically using descriptions in which the distribution of particles in the bulk around the surface is assumed ad hoc to be homogenous, see for example Refs.~\cite{foret2005simple,loose2008,Bement:2015jp}. As a consequence, effective descriptions that only consider the surface are used to analyze the ensuing dynamics. Even though many patterns observed experimentally could be qualitatively reproduced in this way, several observations show that these results have to be taken with care. First, formally the assumption of a homogeneous bulk is not justified. Second, although the patterns may look qualitatively similar to patterns obtained in the full description that also accounts for the bulk, important features may be missed~\cite{Wettmann:2018bp,Levernier.2020}.  

In this work, we consider the case of phase-separating dynamics on a surface in presence of spontaneous exchange of particles between the surface and the bulk, Fig.~\ref{fig:scheme}.
By integrating out the bulk dynamics we obtain an effective model with memory for the time 
evolution on the surface. We show that the resulting dynamics leads to stationary phase-separated patterns with an intrinsic length scale. We determine this length scale both numerically and via variational arguments. We also compare our results to a previous phenomenological model with a simple instantaneous effective kernel~\cite{foret2005simple}. Our work sheds light on the effects of coupling bulk and surface dynamics for pattern formation.

We consider a surface, which coincides with the plane $z=0$, such that we neglect surface fluctuations, exchanging matter (from both sides for simplicity) with a bulk. The system is schematized in  Fig.~\ref{fig:scheme}. We denote by $\vec{r}=(x,y)$ a position on the surface and by $\tilde m(\vec{r})$ and $n(\vec{r},z)$ the densities of particles adhering to the surface and in the bulk respectively. 
Alternatively, we can interpret $\tilde m$ and $n$ as the surface and volume fractions of one component of a two-component system (see Fig.~\ref{fig:scheme}), where the two components could be, for example, protein species that can adhere to a membrane or lipid species that constitute the membrane.
 
Conservation of matter implies that the bulk density $n(\vec{r},z)$ obeys
\begin{equation} \label{eq:n}
\partial_t n + \vec{\nabla} \cdot \vec{j} = 
\delta(z)[k_D \tilde m(\vec{r},t) - k_A n(\vec{r},z=0,t)],
\end{equation}
where $\vec{j}(\vec{r},z,t)=-D\vec{\nabla} n(\vec{r},z,t)$ is the bulk particle current, which we assume to be purely diffusive with diffusion constant $D$. Particles close to the membrane attach to the surface at rate $k_A$, whereas particles on the surface detach at rate $k_D$. We neglect possible cooperative effects during particle attachment and detachment. 

The dynamics of the surface is characterized by either the difference in density of two species
$m = \tilde m_1 - \tilde m_2$ (as shown in Fig.~\ref{fig:scheme}) or by the fluctuations of a single species around some average value $m = \tilde m - \tilde m_0$~\footnote{For a comprehensive explanation of how our model can describe matter distributions of one or two components, please refer to the Supplementary Material}. 
\revthree{We consider a Ginzburg--Landau (GL) free energy, symmetric in $m$, $\F=\int\mathrm{d}^2\vec{r}\left\{-\frac{\alpha}{2}m^2+\frac{\delta}{4}m^4+\frac{\gamma}{2}\left(\nabla m\right)^2\right\}$ with constants $\alpha$, $\beta$, and $\gamma$, to take into account the particles' interactions on the surface. 
This model is interesting due to the non-linearity introduced by the GL term. It applies to the single and the two-species case. The applicability is obvious for the single-component case. In the two-species case it applies when the density fluctuations are much smaller than the average density (see the SM)}. 
For $\alpha>0$, it has two minima at $m_{1,2}=\pm\sqrt{\frac{\alpha}{\delta}}$ showing the tendency for phase separation.
The dynamics of $m(\vec{r},t)$ is ruled by a generalized form of the Cahn-Hilliard equation~\cite{chaikin,cugliandolo_CRP2015_coarsening}:
\begin{equation}\label{eq:m}
\partial_t m + \vec\nabla \cdot [- \mu\vec\nabla\frac{\delta \F}{\delta m}] = k_A n(\vec{r},z=0,t) - k_D m(\vec{r},t)
\end{equation}
where $-\mu \vec\nabla\frac{\delta \F}{\delta m}$ is the surface matter current, $\mu$ a mobility, 
and $n$ denotes in this equation either the difference of the two species $n_1-n_2$ or the 
density in the bulk shifted by $-\frac{k_D}{k_A} m_0$. 

By integrating out the bulk we reduce the two-equations system~\eqref{eq:n} and~\eqref{eq:m} to a single equation for $m(\vec{r},t)$ (see \revtwo{SM}):
\begin{multline}\label{eq:DynamicEvolution}
\frac{\partial m}{\partial t} = - \mu\nabla^2(\alpha m -\delta m^3+\gamma \nabla^2 m) \\
+ \int_0^t dt' \int d\vec{r}'K(\vec{r}-\vec{r}',t-t') m(\vec{r}',t').
\end{multline}
The exchanges of matter with the bulk now manifest in this equation as the kernel $K$
\begin{multline}\label{eq:K(r,t)}
 K(r,t) = {\sqrt{\frac{\pi}{2}}}\frac{k_D\kappa}{{Dt}} e^{-\frac{r^2}{4Dt}}  \Big(\frac{1}{\sqrt{\pi \kappa t}}-e^{\kappa t}\text{Erfc}(\sqrt{\kappa t})\Big) \\
 -k_D\delta(\vec{r}-\vec{r}')\delta(t-t'),
\end{multline}
where ${\kappa}=\pi\frac{k^2_{A}}{2D}$.

The kernel is non-local in time representing a memory in the dynamics coming from the diffusion in the bulk. The result (\ref{eq:DynamicEvolution}) is thus a microscopically rooted description of the dynamics of the surface. It must be compared to more phenomenological approaches \cite{foret2005simple} where the bulk was modelized by a simple ad hoc relaxation term, local in space and time 
\begin{equation}\label{eq:KernelForet}
K_\tau = -\tau^{-1}\delta(\vec{r}-\vec{r}')\delta(t-t'),
\end{equation}
where the parameter $\tau$ is a typical matter exchange time. Then, the dynamics is similar to that of phase separating and reacting chemical mixtures~\cite{Huberman:1976hi,Glotzer:1995ce}. We examine below the physical properties of (\ref{eq:DynamicEvolution}) and show that depending on parameters \revtwo{these} dynamics, which lead to patterns exhibiting characteristic lengthscales, can differ markedly from the phenomenological case. 

\begin{figure}
\begin{center}
 \includegraphics[width=1\linewidth]{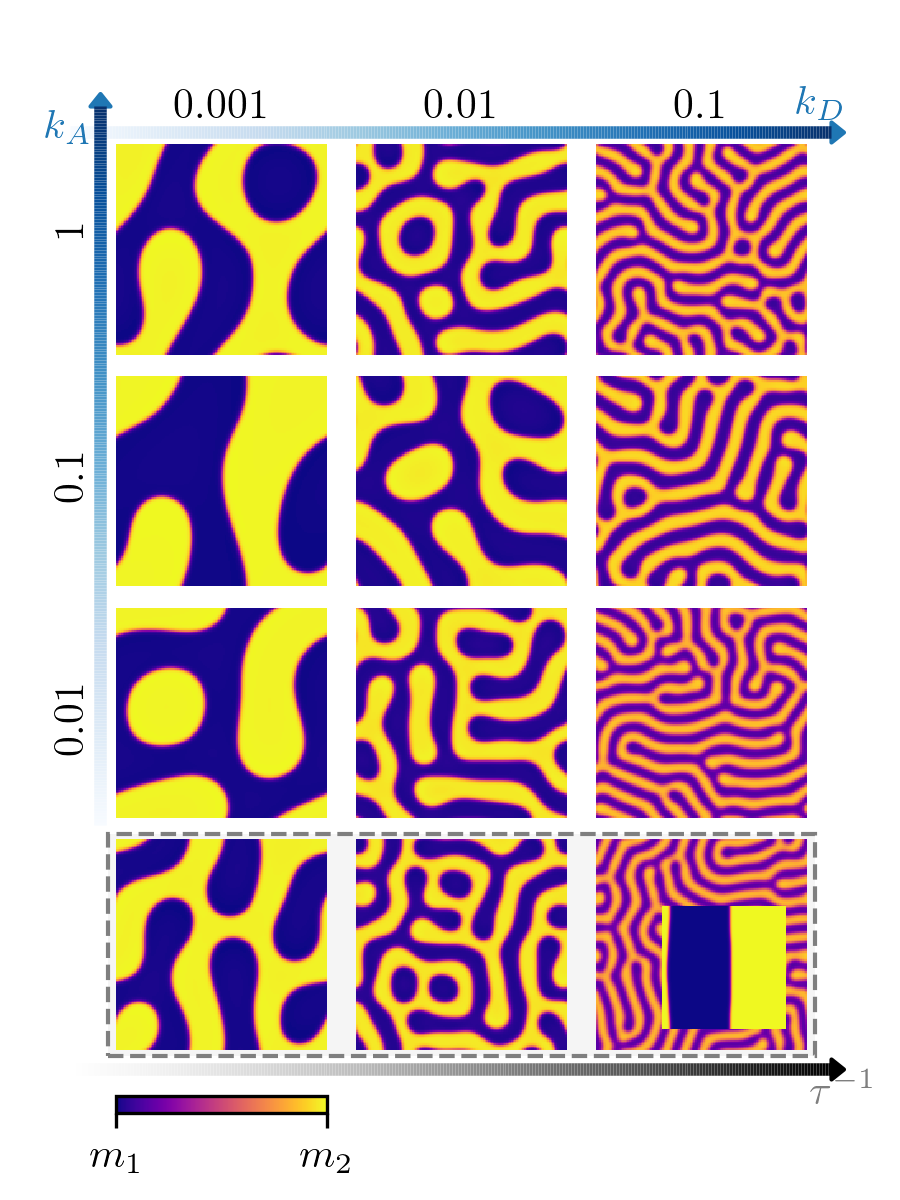}
\end{center}
\caption{\label{fig:snapshots2D}
Snapshots of the distribution $m$ in a domain of 128$\times$128 after $10^6$ simulations steps starting from a random initial condition with $\Sigma_{\vec{r}\in\text{system}}m(\vec{r},t)=0$. The pure states $m_{1}$ and $m_2$ are given by the minima of the Ginzburg--Landau energy. Top rows 1 to 3: configurations for the full kernel \eqref{eq:K(r,t)}. Row 4: configurations for the instantaneous kernel \eqref{eq:KernelForet}. We use $D=0.1$. The values of the attachment rate $k_A$, the detachment rate $k_D$, and the exchange time $\tau$ are given in the figure. In absence of particle exchange with an environment, $K=0$, the system completely phase separates (inset). On the contrary in presence of a kernel $K$ coarsening is limited and a natural scale for the patterns appears.}
\end{figure}

Because of the kernel memory, solving (\ref{eq:DynamicEvolution}) is much more complicated than for the instantaneous kernel (\ref{eq:KernelForet}). To do so we take
advantage of massive parallelization in Graphical Processing Units (GPUs). We consider a system of dimensions $L_x\times L_y$ with $L_x=L_y=128$ unless stated otherwise and with periodic boundary conditions in both directions. We integrate the dynamical equation by using a semi-implicit Fourier-spectral method~\cite{Chen1998_semiimplicitFourierspectral}, adapted from~\cite{caballero_PRE_2018_phi4,caballero_JSTAT_2021_ac}. Without loss of generality, we chose $\mu=\alpha=\delta=\gamma=1$ and space discretization equal to 1. We use $D$ to fix the time-scale by choosing $D=0.1$ and vary $k_A$ and $k_D$. We approximate the integral in (\ref{eq:DynamicEvolution}) by a Riemann sum that requires the configurations of the system $m(\vec{q},t)$ in the previous $M$ simulation steps (see SM for more details). For an integration time-step $\Delta t=10^{-1}$ the difference in $m(\vec{r},t)$ for simulations with $M=100$ and $M=10$ is of the order of $10^{-2}$. For our purpose, this is a reasonable numerical error so we fix $\Delta t=10^{-1}$ and $M=10$.
Typical results obtained from a random initial condition are shown in Fig.~\ref{fig:snapshots2D} together with results for the phenomenological kernel (\ref{eq:KernelForet}) for several values of $\tau$. 

In absence of matter exchange with the bath, $K=0$, and Eq.~(\ref{eq:DynamicEvolution}) reduces to the Cahn-Hilliard equation for which coarsening leads to macroscopic phase separation with eventually two domains of the pure phases $m_1$ and $m_2$. Starting from a random initial condition with  $\Sigma_{\vec{r}\in\text{system}}m(\vec{r},t)=0$, we observe macroscopic separation after $\sim 10^5$ simulation steps, see insert of Fig.~\ref{fig:snapshots2D}.

On the contrary, in the presence of the full kernel \eqref{eq:K(r,t)} coarsening is interrupted and a natural length scale of the pattern emerges. 
This can be seen in Fig.~\ref{fig:snapshots2D}, where configurations, which have evolved from the same random initial condition, but for a much longer time ($10^6$ steps), exhibit a characteristic pattern. \revtwo{Membrane-bound particles in steady-state exhibit a current which vanishes in absence of matter exchange with the bulk (see SM)}.
To ensure that we were not tricked by slowing down of the dynamics towards complete phase separation, we also considered initial conditions with fully separated phases. In the presence of the full kernel \eqref{eq:K(r,t)}, stripe or bubble configurations evolved into multiple domains, indicating that patterns with a characteristic length scale are indeed stable fixed points of the dynamics, see SM. \nccom{In addition to the labyrinthine patterns shown in Fig.~\ref{fig:snapshots2D}, which resemble patterns observed in the \textit{Escherichia coli} Min system~\cite{Glock_2018}, after a shift in the potential we also found circular patterns, see SM, corresponding to protein or lipid domains frequently found in cells. For simplicity, we continue in the following with the non-shifted potential, but our analysis is readily applicable also in the shifted case.}

\nccom{The steady state patterns} exhibit a characteristic length scale determining the width of the meandering stripes. This length scale decreases with increasing detachment rate $k_D$ and shows a non-monotonic dependence on the attachment rate $k_A$. In the case of the instantaneous kernel (\ref{eq:KernelForet}) we observe a similar dependence of the characteristic length scale on $\tau^{-1}$. \nccom{We can estimate the scales of these domains by introducing units to our numerical simulations. For the protein MinD in \textit{E. coli}, the residence time on the membrane was measured \textit{in vitro} to be of the order of $10s$ ($k^{MinD}_D\simeq 10^{-1}s^{-1}$) and its diffusion constant $D^{MinD}$ of the order $10^{-1}\mu m^2/s$~\cite{loose2011min}. For example, for $k_D=10^{-2}$ we obtain a time scale $t_0=\frac{k_D}{k^{MinD}_D}\simeq10^{-1}s$. Since in our simulations we use $D=0.1$ this sets the length units to $\xi_0=\sqrt{\frac{D^{MinD}}{D}t_0}\simeq \frac{1}{3}\mu m$. The attachment rate depends on the cytosolic protein concentration and is more difficult to get. In particular, there might be cooperative effects, such that the attachment rate can depend on the amount of proteins on the membrane. For MinD \textit{in vitro} a rate $10^{-3}\mu m/s$ -with a buffer density of $1000 \mu m^{-3}$- has been previously used in simulations. This gives $k_A\simeq \frac{1}{3}10^{-3}$. For these values we get domains sizes of approximately $20 \xi_0\simeq 10\mu m$, which are of the order of observed domains~\cite{Glock_2018}.}

In order to rationalize \nccom{the} dependence of the patterns on the matter exchange rates and to quantify the differences between the full and the instantaneous kernels, we first examine the full kernel as a function of momentum and the parameter $s$ resulting from a Laplace transform of the temporal coordinate. In these variables, our kernel (\ref{eq:K(r,t)}) reads
\begin{equation}
 K(q,s) = -\frac{k_D}{1+\sqrt{\frac{ \kappa }{Dq^2+s}}}. 
\label{eq:K(q,s)}
\end{equation}
We see that for small values of the parameter $\kappa$, that is small values of the attachment rate $k_A$ or large diffusion constants $D$, this expression becomes essentially independent of $s$ and thus an instantaneous kernel of the form (\ref{eq:KernelForet}) with the identification $\tau = 1/k_D$. Our microscopic calculation thus validates the use of the instantaneous kernel (\ref{eq:KernelForet}) in such a limit and gives a microscopic value for the effective lifetime $\tau$. In the opposite limit, on the contrary, we see that the non local dependence on time has a strong effect on the kernel and we can thus expect different physical behaviors, at least quantitatively. 


\begin{figure}
\begin{center}
 \includegraphics[width=1\linewidth]{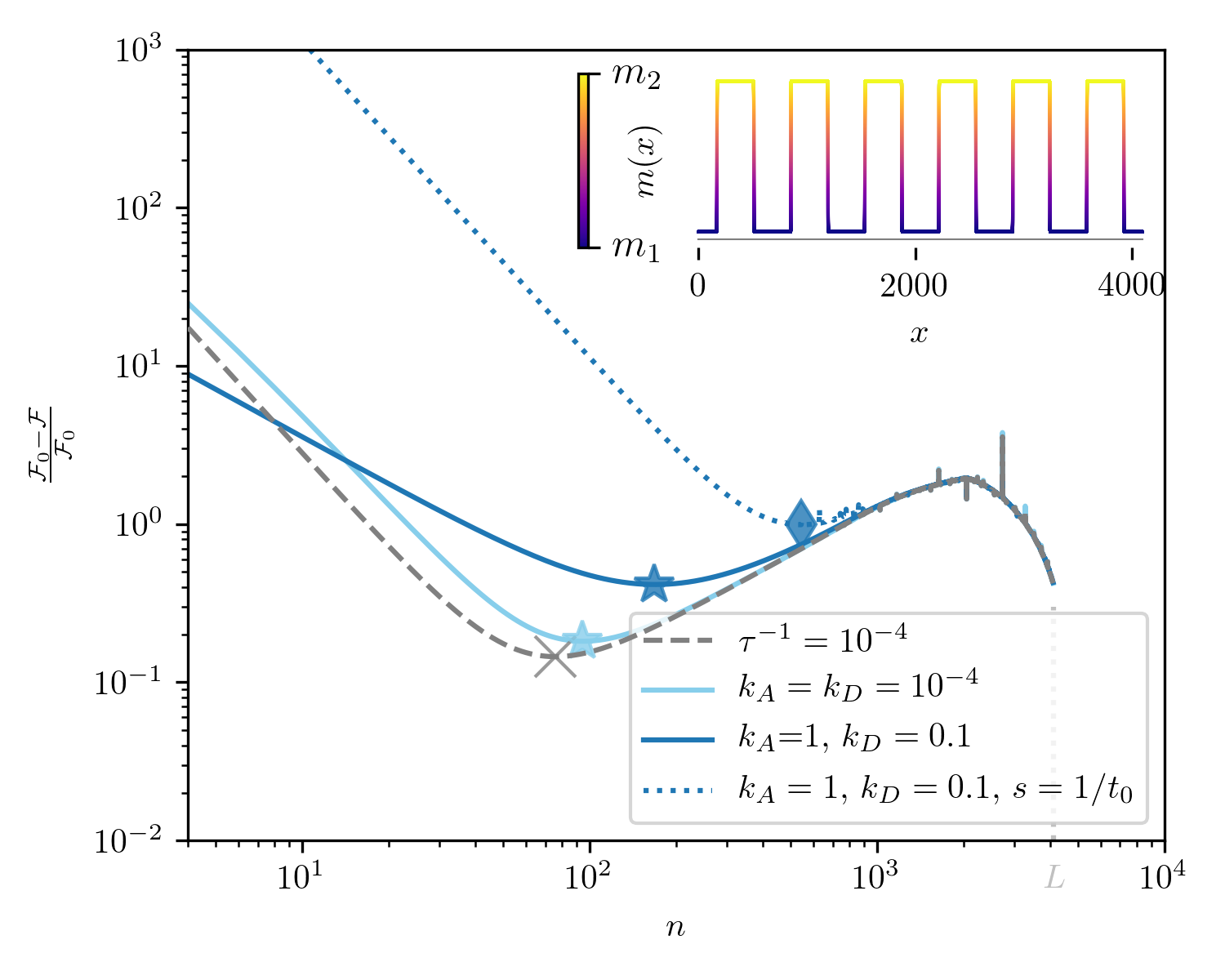}
\end{center} 
 \caption{\label{fig:energy_profiles} Pseudo free energy $\F_m$ for profiles with $n$ evenly spaced kinks. Dashed line: instantaneous kernel, full lines: full kernel, dotted line: full kernel for $s=1/t_0=1$. Kernel parameters $\tau^{-1}$, $k_D$, and $k_A$ as indicated in the legend. The minimum $n_{min}$ of each energy function is indicated with a symbol. Inset: profile with $n=12$ kinks.}
\end{figure}

Let us consider the instantaneous kernel (\ref{eq:KernelForet}) in one spatial dimension. Its dynamics is given by 
\begin{align} \label{eq:ForetDyanmicsFourier}
 \partial_t m(q,t)&= -q^2 \mu\frac{\delta \F_m}{\delta m^*({q},t)},    
\end{align}
where the star indicates the complex conjugate, and $q=\frac{2\pi}{L}k$ for a system size $L$, and $k=1,\dots, L$. In this expression, we have introduced the pseudo-free energy $\F_m = \F + \F_\tau$, where $\F$ is the GL free energy and $\F_\tau$ satisfies
\begin{equation} \label{eq:ftau}
 \frac{\delta \F_\tau}{\delta m^*({q},t)} = \frac{1}{\tau q^2 \mu} m({q},t),    
\end{equation}
From Eq.~(\ref{eq:ForetDyanmicsFourier}) it is easy to show that $\F_m$ monotonically decreases under the time evolution and thus that the fixed point of the dynamic evolution must correspond to the minimum of $\F_m$ if reachable from the initial configuration.
For the parameters of the GL free energy used in this work, we can approximate these states as regions of uniform concentration, where $m$ takes one of the minimal values $m_1$ or $m_2$, separated by narrow transition regions or 'kinks'. 
In the case of the Cahn-Hilliard equation, where $\F_\tau=0$, these kinks take the form of a hyperbolic tangent, $\sqrt{\frac{\alpha}{\delta}}\tanh(
\sqrt{\frac{\alpha}{2\gamma}}x)$, and $\F\approx\F_0+\epsilon_0 n$. Here, $n$ is the (even) number of kinks, $\epsilon_0$ the energy associated with a kink in GL, and $\F_0$ the energy associated with the uniform regions. The term $\F_0$ depends only weakly on $n$. The configuration with minimal energy is thus the one with the minimal number of kinks, i.e., $n=2$, and corresponds to macroscopic phase separation. 

The presence of $\F_\tau$ changes this minimum. We can estimate the corresponding number of domains through a variational approach. First, we construct one-dimensional profiles by combining $n$ evenly spaced kinks and probe the value of the pseudo-free energy as a function of $n$ as shown in Fig.~\ref{fig:energy_profiles}. For our  parameters, kink (and anti-kink) $j$ with $j=0,\cdots, n/2$ is given by $\theta \tanh(\frac{x-x_j(\theta)}{\sqrt{2}})$, where $\theta=1$ for $x\in [2jL/n,(2j+1)L/n]$, $\theta=-1$ for $x\in[(2j+1)L/n,2(j+1)L/n]$ and $x_j(\theta)=(4j+3/2-\theta/2)L/(2n)$. See Fig.~\ref{fig:energy_profiles}, inset for an example of the resulting profile. Then, we compute the pseudo-free energy $\F_m$ for this profile and minimize with respect to $n$.

Extending this analysis to the case of the full kernel \eqref{eq:K(r,t)} is more involved. A naive attempt could be made from Eq.~(\ref{eq:K(q,s)}) by assuming that at large times we can approximate this expression by taking $s\to 0$ and then identifying the resulting $q$-dependent prefactor with $1/\tau$ in Eq.~(\ref{eq:ftau}). The pseudo-free energy has in this case a qualitatively similar shape as for the instantaneous kernel, Fig.~\ref{fig:energy_profiles}. 

\begin{figure}
 \begin{center}
 \includegraphics[width=1\linewidth]{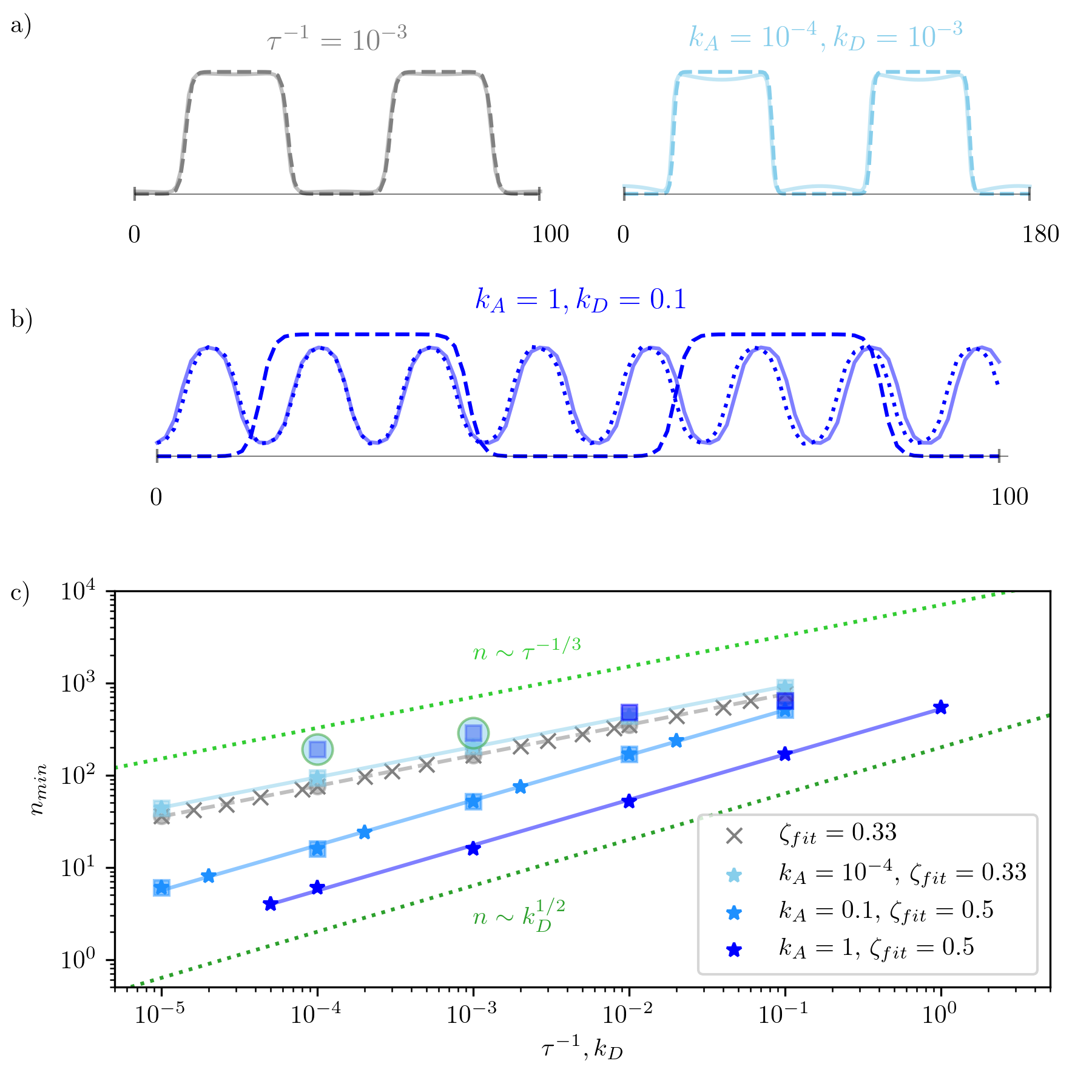}
 \end{center}
 \caption{\label{fig:n} a) Part of the profiles used as initial condition (dashed lines) to simulate systems evolving with (\ref{eq:KernelForet}) ($\tau^{-1}=10^{-4}$) and our kernel (\ref{eq:K(r,t)}) for $k_A=10^{-4}$ and $k_D=\tau^{-1}$. In continuous lines we show the profiles after dynamic evolution. For $k_A=1$ the number of kinks significantly differs from the number used as initial condition, as shown in b). For reference in this case we show in dotted lines a constructed profile with the observed $n_{min}$ and amplitude 0.8. c) Scaling relation between the optimal number of kinks $n_{min}$ and the kernel parameters $k_D$ for fixed values of $k_A$ or $\tau^{-1}$ depending on the kernel used. Stars and crosses represent respectively for the microscopic and phenomenological kernel an initial condition obtained by minimizing the effective free energy, as shown in Fig.~\ref{fig:energy_profiles}. Squares indicate the number of kinks observed in simulations of systems that evolved for up to $10^6$ steps from this initial condition and thus correspond up to numerical limitations to the fixed point of the dynamics (the two circles corresponding to $k_A=1$ and $k_D=10^{-4}, 10^{-3}$ indicate that the configuration observed for these set of parameters is not regular in contrast with all the other observed configurations). The scaling of $n_{min}$ appears to be always a power law but with strong quantitative differences between the microscopic and phenomenological kernels as expected when $k_A$ is large. Fits of the data with a power law with exponent $\zeta_{dis}$ are shown in dashed and continuous lines for their respective kernels.}
\end{figure}

We test the configurations obtained with the minimization strategy by constructing profiles with $n_{min}$ number of kinks that serve as initial condition for the full evolution of the dynamic Eq.~(\ref{eq:DynamicEvolution}) for both kernels. As shown in Fig.~\ref{fig:n}a, the effect of both kernels on the profiles is to slightly distort the shape of the valleys and peaks. As depicted in Fig.~\ref{fig:n}b we see that for the instantaneous kernel, the variational solution is essentially stable under time evolution, showing that the variational principle is indeed predicting correctly the fixed point of the dynamics. A simple estimate can be given by noting that $\F_\tau$ scales as $L^2\tau^{-1}n^{-2}$. In presence of matter exchange in the instantaneous system, we find that the minimum of $\F_m$ is reached for $n=const \left(\frac{L^2}{\tau\epsilon_0}\right)^{1/3}$. This scaling relation is in agreement with our numerical results as shown in Fig.~\ref{fig:n}c. 
However, for the full kernel, the final state differs strongly from the initial configuration derived from the argument above when $k_A$ is large. In particular, the number of kinks is largely different between the variational estimate and the full evolution, Fig.~\ref{fig:n}b.
This shows that the naive  substitution of the $s=0$-kernel in the pseudo-free energy is not sufficient and that a more precise method must be found. Putting phenomenologically a finite $s$, as shown in Fig.~\ref{fig:energy_profiles}, to mimic a finite time cutoff in the memory of the full kernel does push the minimum of the pseudo free-energy to a larger number of kinks but does not allow 
for a reliable prediction of the fixed point of the time evolution. Another possibility is that there is more than just one characteristic scale for the domains which could explain why deformed bubble domains were observed in liquid-liquid phase separation of intracellular condensates~\cite{riback2020composition}.
Finding the equivalent of a predictive variational approach, if at all possible, for the full kernel is a very interesting but challenging question for future studies. Extension of these methods to the case of the two dimensional patterns computed numerically in Fig.~\ref{fig:snapshots2D} is also interesting since it would provide direct access to the pattern formation bypassing the need for the full dynamical analysis.

Beyond the analytical approximations, our simulations show clearly that an optimal length scale exists, revealed by the optimal number of kinks $n_{min}$. This number scaling algebraically with the parameters of the kernel. As expected the full and instantaneous kernels essentially coincide at small $k_A$, whereas largely different behaviors with different exponents are observed when $k_A$ is large; at least for the time available in our simulations. This conveys the importance of having a properly defined full kernel to identify quantitatively the pattern formed. 

In this work, we have presented a field theory to describe matter distribution in a membrane exchanging matter with its environment. Our theory predicts arrested phase separation with domains characterized by typical sizes determined by the absorption and expulsion matter exchange rates. Based on semi-microscopical equations for the composition of a surface and its environment, we integrate the environment contribution in a single equation for the surface composition dynamics. The matter exchange effect induces spatio-temporal memory effects with non-trivial consequences for the typical domain sizes for large absorption rates. On the contrary, when the absorption rate is low ($k_A\ll 1$) our theory behaves very similarly to an instantaneous kernel that was previously phenomenologically proposed. In this case, we show with semi-analytical arguments that matter exchange induces phase separation in the membrane with domains characterized by a typical length. We compute its scaling as a function of the parameters of the problem for the one-dimensional case. Our theory provides a physical justification for the functional form of the instantaneous kernel.

\nccom{Our theory shows that when the diffusion constant is large or the adsorption rate is low particles detached from the membrane are relatively quickly reabsorbed and homogenized in the bath. The field $n$ describing matter distribution in the environment thus does not play any role and can be neglected in Eq.~\ref{eq:m}. In this case, the instantaneous kernel can capture the physics of matter exchange on the surface. However, for low diffusion constants or large adsorption rates, the opposite happens and the membrane 'remembers' the previous states of the particles. In this case, the instantaneous kernel fails to capture the physics of the problem and the full kernel should be considered.} \revtwo{In this case, predicting the domain scaling behaviour is more involved and requires further investigation (see the current behaviour in the SM).}

\nccom{In future work it will be interesting to study the interplay between phase separation as discussed above and interactions between different lipids and/or proteins induced by membrane undulations. Indeed, such undulations have been argued to induce interactions between transmembrane proteins~\cite{Goulian.1993,Park:1996} and different lipids in the same~\cite{Dean.2015} or in opposite leaflets of the bilayer membrane~\cite{Haataja.2017} and, for large distances, exceed van der Waals or electrostatic forces. As these forces can be attractive or repulsive, we expect a large number of phases to be generated in this case.}

\begin{acknowledgments}
This work was supported in part by the Swiss National Science Foundation under Division II (grant 200020-188687).
All numerical simulations were performed at the University of Geneva on the \textit{Mafalda} cluster of GPUs. 
\end{acknowledgments}

\typeout{}


%

\clearpage

\onecolumngrid
\begin{center}
\medskip
\begin{large}
\textbf{Phase separation in surfaces in presence of matter exchange\\
\textit{Supplementary material}}
\end{large}
\end{center}

\bigskip

\twocolumngrid
\setcounter{equation}{0}
\renewcommand{\theequation}{S.\arabic{equation}}

\setcounter{figure}{0}
\renewcommand{\thefigure}{S.\arabic{figure}}

\section*{Model for one or two components}
\label{app:two-components}

As discussed below our model can describe both, matter distributions of one or two components. 

\subsection{\revfour{One component}}

Consider first the case of one particle species ($I$). The governing equations are
\begin{eqnarray}
\nonumber 
\partial_t n_i - D {\nabla}^2 n_i &=&
\delta(z)[k_D m_i(\vec{r},t) - k_A n_i(\vec{r},z=0,t)]\\
\partial_t m_i + \vec\nabla \cdot \vec{j}_{m_i} &=& k_A n_i(\vec{r},z=0,t) - k_D m_i(\vec{r},t),
\label{eq:explicitModel}
\end{eqnarray}
where $i=I$, $m_I$ and $n_I$ are the particle densities on the membrane and in the bulk, respectively. The constants $k_A$ and $k_D$ denote the attachment and detachment rates of particles to and from the membrane, $D$ is the bulk diffusion constant and $\vec j_{m_I}$ is the particle current on the membrane.
We write $n_I=n_0+n$ and $m_I=m_0+ m$, where $n_0$ is the particle density infinitely far away from the membrane and $m_0=k_An_0/k_D$. The equations for $n$ and $m$ are then the same as Eqs.~(1) and (2). 
\revthree{Note that when the ratio $k_A/k_D$ is varied one needs to maintain the average density on the surface constant by adjusting the density $n_0$ in the bulk accordingly so that the Ginzburg-Landau (GL) free energy remains symmetrical in $m$. This can always be done except in singular cases such as $k_A=0$.}

The values of $m$ are restricted to $m\ge -m_0$ for the density of membrane-bound particles to be positive. For the values of $\alpha$ and $\delta$ of the GL energy used in the manuscript, we have minima at $m_{1,2}=\mp1$. Our simulations show that the values of the density $m$ do not exceed the interval $[m_1,m_2]$ such that for $m_0\ge1$ the density of membrane-bound particles is positive as required.

\subsection{\revfour{Two components}}

\revtwo{For the two species case, the four governing equations for particles of the two types can be written as in Eq.~(\ref{eq:explicitModel}) by taking $i=I, II$. In this case, $m_{i}$ and $n_{i}$ are the particle densities of type $i=I, II$ on the membrane and in the bulk, respectively. The constants $k_{A}$ and $k_{D}$ denote the attachment and detachment rates of particles $i=I, II$ to and from the membrane. We assume that both particle types attach and detach with the same rates $k_{AI}=k_{AII}=k_A$ and $k_{DI}=k_{DII}=k_D$. Furthermore, we set
$\vec{j}_m\equiv\vec{j}_I-\vec{j}_{II}=-\mu\vec\nabla\frac{\delta \mathcal{F}}{\delta m} $, such that the dynamic equations for $n=n_I-n_{II}$ and $m=m_I-m_{II}$ are again of the form of Eqs.~(1) and (2) of the main text.}

\revfour{The GL form depending on the density differences can be used in the limit when the fluctuations of the difference in density are small compared to the total density on the membrane. This can be done in two ways: either by considering a "three-state problem" or by assuming a constrained density on the membrane.}

\revfour{\textbf{GL energy for a ``three state'' problem}. Let us first consider two independent species. The linear terms can be added and these depend only on the density difference. The main question is whether one can write a GL term that would also depend only on the density difference, knowing that now the total density $n_1 + n_2$ can fluctuate on the membrane given the independence of the two species.}

\revfour{The answer can be obtained by looking at e.g. a spin one model where the three states $\sigma_i = \pm 1$ would represent species $I$ and $II$ and the state $\sigma_i=0$ would be an empty site. A phenomenological Hamiltonian accounting for the essential features of this situation is 
\begin{equation}
    H = - J \sum_{(i,j)} \sigma_i \sigma_j - D \sum_j \sigma_j^2, 
\end{equation}
where $(i,j)$ denotes nearest neighbors on some lattice and $D$ is a parameter controlling the proportion of ``occupied'' versus ``empty'' sites. We consider $J >0$ which favours particles of the same species being close to each other. One can derive the GL expression for this Hamiltonian by Feynman's variational approach~\cite{feynman2018statistical}.}

\revfour{Let us denote by 
\begin{equation}
    T_i = \langle \sigma_i^2 \rangle
\end{equation}
the density of occupied sites. The difference between the two species is given by 
\begin{equation}
    m_i = \langle \sigma_i \rangle.
\end{equation}
In the mean-field limit, one has for $m_i = 0$
\begin{equation}
    T_i = \frac{2 e^{D/T}}{1 + 2 e^{D/T}}.
\end{equation}
Small deviations from this state can be parametrized by $m_i$ and $x_i$ with 
\begin{equation}
    T_i = \frac{2 e^{D/T}}{1 + 2 e^{D/T}} + x_i.
\end{equation}
A simultaneous expansion in $m_i$ and $x_i$ gives the free energy
\begin{widetext}
\begin{equation}
\label{eq:PottsFE}
\begin{split}
    \Gamma &= \left[-T \log \left(2 e^{\frac{D}{T}}+1\right)+\frac{1}{4} T x^2 \left(3 \sinh \frac{D}{T}+5 \cosh \frac{D}{T}+4\right)+O\left(x^3\right)\right] \\
    & + m^2 \left[\frac{1}{4} \left(2+e^{-\frac{D}{T}}\right) T-J+\frac{1}{8} \left(-e^{-\frac{2 D}{T}} \left(1+2 e^{\frac{D}{T}}\right)^2\right) T x
    +O\left(x^2\right)\right] \\
    & +m^4 \left[\frac{1}{96} \left(e^{-\frac{3 D}{T}} \left(1+2 e^{\frac{D}{T}}\right)^3\right) T+O\left(x\right)\right]+O\left(m^5\right)
\end{split}
\end{equation}
\end{widetext}}

\revfour{For large values of $D$, the coefficient of the $x^2$ term in Eq.~\eqref{eq:PottsFE} is positive and large and thus essentially imposes $x=0$ i.e. the total density is essentially frozen to its average value. In the same limit, the coefficients of $m^2$ and $m^4$ terms converge to finite values, whereas the term $m^2 x$ is negligible compared to the $m^2$ term (except when extremely close to the transition point). We thus recover the standard GL expansion in terms of the density difference.}


\revfour{\textbf{Constrained density on the membrane}. As a second case, let us consider the case where the lipids of the membrane themselves exchange with the environment. Let us assume that the (fluid) lipid membrane consists of two kinds of lipids. The total two-dimensional lipid membrane density $m_\mathrm{tot}$ is constant, and we denote the densities of lipids in the environment by $n_I$ and $n_{II}$ with $n_\mathrm{tot}=n_I+n_{II}$ being the total lipid bulk density. Lipid molecules can leave the membrane and new ones can go in. Since $m_\mathrm{tot}=const$, each leaving lipid is immediately replaced by another lipid molecule from the bulk. In this case, the free energy can be expressed solely in terms of the membrane density $m$ of lipids of type I.}

\revfour{Let $k_d$ denote the rate at which lipid molecules of both kinds leave the membrane. Changes in the density $m$ due to the exchange of lipids with the bulk evolve according to
\[
\dot{m} = -k_d m\frac{n_{II}}{n_\mathrm{tot}} +k_d(m_\mathrm{tot}-m)\frac{n_I}{n_\mathrm{tot}}.
\]
This rate of change can be expressed in terms of the densities of lipid $I$ only
\begin{align}
\dot{m}&=-k_dm\frac{n_\mathrm{tot}-n_{I}}{n_\mathrm{tot}}+k_d(m_\mathrm{tot}-m)\frac{n_{I}}{n_\mathrm{tot}}\nonumber\\
&\equiv -k_dm+k_an_I,
\label{eq:final}
\end{align}
where $k_a = k_dm_\mathrm{tot}/n_\mathrm{tot}$. 
Since $n_\mathrm{tot}=const$, Eq.~\eqref{eq:final} has the same form used in the main article.} 



\section*{Reduction of the coupled dynamic equations}
\label{app:equations-reduction}

The system of two equations for the surface membrane and the reservoir
\begin{equation}
\systeme[][:]{\partial_t n  = D \Delta n+\delta(z)[k_D m - k_An(\vec{r},z=0,t)]:
\partial_t m= \mu\nabla^2\frac{\delta \F}{\delta m}+k_A n(\vec{r},z=0,t) - k_Dm},\\
\label{eq:coupledBath-membrane}
\end{equation}
can be reduced to a single equation. By defining the Fourier transform of a function $A_x$ as $A_{x}=\int \frac{d{q}}{\sqrt{2\pi}}e^{i{x}{q}}\tilde A_{q}$, we write
\begin{equation}
  \tilde n(\vec{q},z=0,w) = \frac{k_D \tilde m(\vec{q},w)\tilde I(\vec{q},w)}{\sqrt{2\pi}+k_A\tilde I(\vec{q},w)}, \\
  \label{eq:nq0w}
\end{equation}
where
$\tilde I(\vec{q},w)=\int dk_z\frac{1}{iw +D(k_z^2+\vec{q}^2)}$. Replacing (\ref{eq:nq0w}) in (\ref{eq:coupledBath-membrane}) for $\tilde I\neq 0$ we obtain
\[
\partial_t m= \mu\nabla^2\frac{\delta \F}{\delta m}+\F^{-1}\Big[\Big(\tilde K(\vec{q},w) - 1\Big)k_D\tilde m(\vec{q},w))\Big],
\] with $
\tilde K(\vec{q}, w)=\frac{1}{\frac{\sqrt{2\pi}}{\tilde I(\vec{q},w)}+k_A }.$

The \revtwo{time}-Fourier transform of this quantity is
\begin{equation}
\hat K(\vec{q},t)=\kappa e^{-q^2Dt}\Big(\frac{1}{\sqrt{\pi \kappa t}}-\text{Erfcx}(\sqrt{\kappa t})\Big),\\
\label{eq:K(q,t)}
\end{equation}
where ${\kappa}=\pi\frac{k^2_{A}}{2D}$, and $\text{Erfcx}(t)$ is the standard error function $\text{Erfcx}(t)=e^{t^2}\frac{2}{\sqrt{\pi}}\int_t^\infty e^{-z^2}dz$.

\section*{Numerical solution}

We discretize Eq. 3 so that the surface particles' density $m$ is a matrix of elements $(i,j)$ at each simulation step $p$, with a given initial condition $m(i,j,p=0)$. To obtain the evolution of $m$ we use the standard Euler semi-implicit integration method for the time variable and a five-point discretization of the Laplacian terms in Fourier space for the spatial coordinates. 

Compared to a standard Cahn-Hilliard equation we have to consider an extra term involving the integral of our kernel. We approximate the integral in Eq. 3 at a simulation step $p$ by the Riemann sum
$\int_{0}^{t}dt'\hat K(\vec{q},t-t')m(\vec{q},t')\simeq \sum_{p'=M}^{p-1}\hat K(q_i,q_j,p-p')m(q_i,q_j,p')\Delta t$, where $\hat K$ is given by Eq.~\ref{eq:K(q,t)}, and $M$ is the number of previous configurations that will be taken into account in the calculation. 

\begin{figure}
\begin{center}
{\includegraphics[width=1\linewidth]{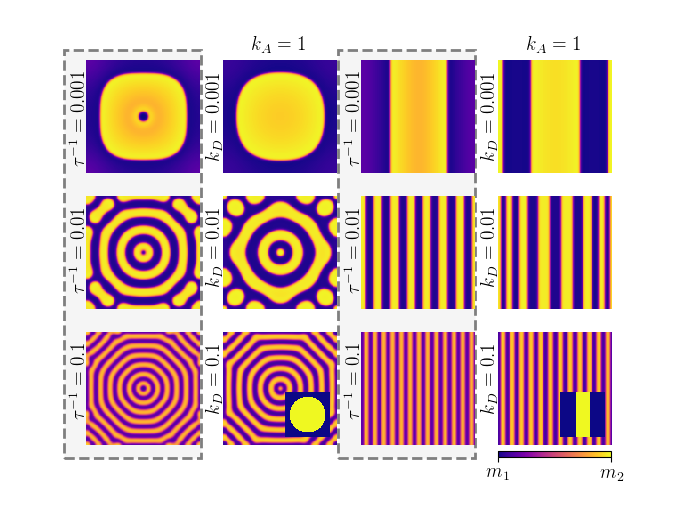}}
\end{center}
\caption{Snapshot of systems that evolved during $10^4$ simulation steps for different values of the absorption and desorption rates, $k_A$ and $k_D$, respectively, at fixed diffusion constant $D = 0.1$. We consider two different initial conditions (shown on the insets). The effect of matter exchange between the membrane and the bath is observed
at relatively short times, especially for larger values of $k_D$. The images highlighted with dashed grey lines correspond to systems that evolved under the instantaneous kernel (Eq. 5).}
  \label{fig:initial_conditions}
\end{figure}

\section*{Short-time effects}

To further verify the multi-domain state of the system when matter exchange is considered, we studied
the cases where the initial condition is a single domain, either a bubble or a stripe, as shown in Fig.~\ref{fig:initial_conditions}. The effect of the kernel is immediately observed in the simulations for both considered kernels.

\section*{Shifted potential}

\begin{figure}
\begin{center}
{\includegraphics[width=1\linewidth]{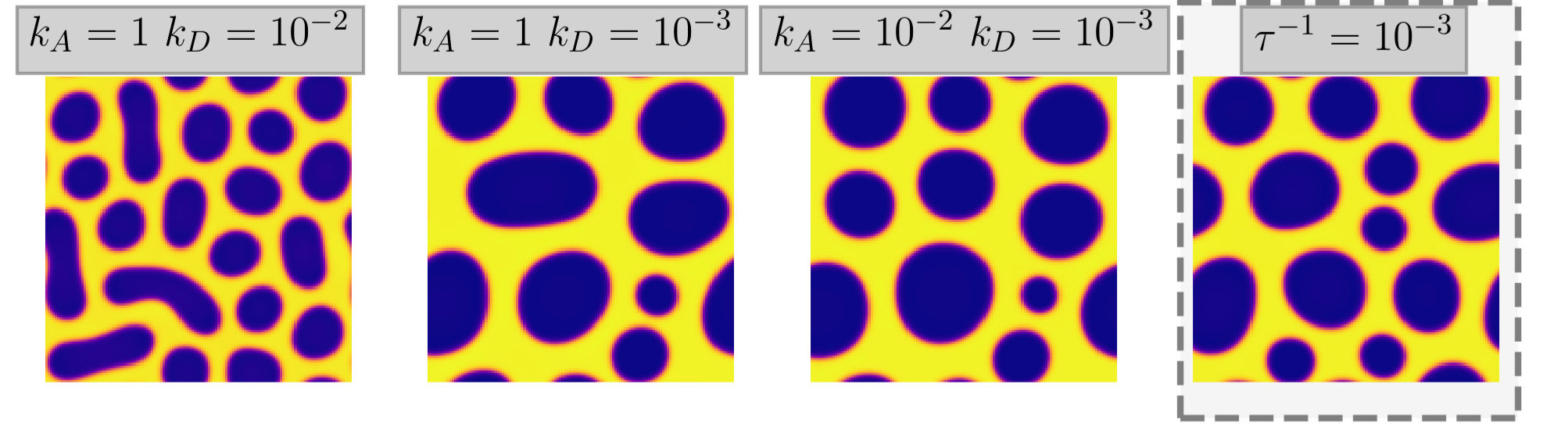}}
\end{center}
\caption{\nccom{Simulations were run under the same conditions as in Fig. 2 of the main text: we start from a random initial condition and let the system evolve for $10^5$ steps. We now consider a potential shift $m_c=0.2$.}}
  \label{fig:bubbles}
\end{figure}

\nccom{The Ginzburg-Landau free energy can be shifted $\F=\int\mathrm{d}^2\vec{r}\left\{-\frac{\alpha}{2}(m-m_c)^2+\frac{\delta}{4}(m-m_c)^4+\frac{\gamma}{2}\left(\nabla m\right)^2\right\}$ -as done, for example in~\cite{foret2005simple}. \revthree{This implies that the initial matter distribution in the membrane is shifted towards one minimum by $m_c$.} As a result, the minima of the double-well potential are shifted towards larger values of $m$ \revthree{and tilted to favour one minimum. W}e observe a bubble-like distribution of domains as shown in Fig.~\ref{fig:bubbles}}.

\section*{Membrane current}

\begin{figure}
\begin{center}
{\includegraphics[width=1.1\linewidth]{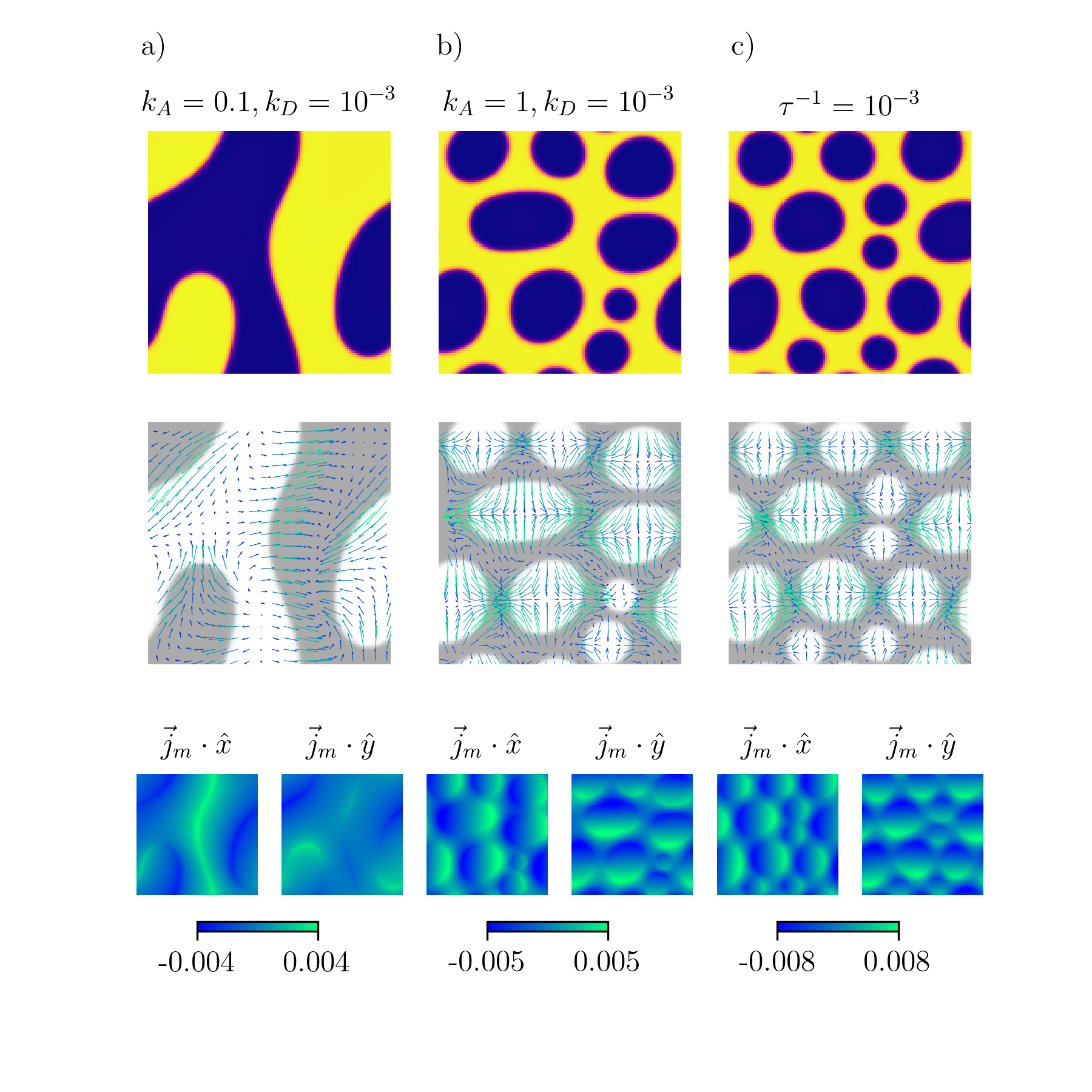}}
\end{center}
\caption{\revtwo{Top: steady-state configurations observed for a) $k_A=0.1$ and $k_D=10^{-3}$ after $10^6$ simulation steps (same as that shown in Figure 2 of our manuscript) b) and c) $k_A=1$, $k_D=10^{-3}$ and $\tau^{-1}=10^{-3}$, respectively, both with shifted potential $m_c=0.2$ after $10^5$ simulation steps. Central: surface matter current $\mathbf{j}_m=-\mu\vec\nabla\frac{\delta  \mathcal{F}}{\delta m}$ obtained for the three different configurations. Bottom: surface matter current along the $\hat{x}$ and $\hat{y}$-directions.}}
  \label{fig:current}
\end{figure}

\revtwo{The non-equilibrium character of our system is clearly expressed through the presence of a current of membrane-bound particles in a steady state, see Figure~\ref{fig:current}. Due to this current, particles attach to and detach from the membrane at different locations. For $k_a=k_D=0$, the current vanishes and the system eventually settles into an equilibrium state that is globally phase-separated. 
If one could write the velocity of the membrane-bound particles $\vec{v}$, the current of membrane-bound particles $\vec{j}_m$ written as $\vec{j}_m=\vec{v} m$ would give the length scale of the domains through $v/k_D$.
}

\end{document}